\documentclass[11pt, a4paper]{article}
\usepackage{jheppub}
\pdfoutput=1

\begin{document}

\newcommand{\be}{\begin{equation}}
\newcommand{\ee}{\end{equation}}
\newcommand{\bea}{\begin{eqnarray}}
\newcommand{\eea}{\end{eqnarray}}
\newcommand{\ba}{\begin{array}}
\newcommand{\ea}{\end{array}}
\newcommand{\pt}{\partial}
\newcommand{\nn}{\nonumber}
\def\tr{\mathop{\rm tr}\nolimits}
\def\Tr{\mathop{\rm Tr}\nolimits}
\newcommand{\dsf}{\displaystyle\frac}
\newcommand{\ds}{\displaystyle}
\newcommand{\pz}{\partial_z}
\newcommand{\dz}{\Delta_z}
\newcommand{\dzf}{\Delta_{z,4}}
\newcommand{\ypp}{f^{\prime\prime}}
\newcommand{\yp}{f^{\prime}}
\newcommand{\Zpp}{Z^{\prime\prime}}
\newcommand{\Zp}{Z^{\prime}}
\newcommand{\dzp}{\Delta^+}
\newcommand{\dzm}{\Delta^-}
\newcommand{\dzpt}{\tilde\Delta^+}
\newcommand{\dzmt}{\tilde\Delta^-}
\newcommand{\dzpm}{\Delta^\pm}
\newcommand{\dzmp}{\Delta^\mp}
\newcommand{\mn}{p}
\newcommand{\mnm}{p_m}
\newcommand{\tM}{\tilde M}
\newcommand{\ts}{\tilde\sigma}
\newcommand{\tp}{\tilde\mn}


\def\a{\alpha}
\def\b{\beta}
\def\c{\gamma}
\def\d{\delta}
\def\e{\epsilon}
\def\ve{\varepsilon}
\def\f{\phi}
\def\v{\varphi}
\def\g{\gamma}
\def\h{\eta}
\def\i{\iota}
\def\j{\psi}
\def\k{\kappa}
\def\l{\lambda}
\def\m{\mu}
\def\n{\nu}
\def\o{\omega}
\def\p{\pi}
\def\th{\theta}
\def\r{\rho}
\def\s{\sigma}
\def\t{\tau}
\def\u{\upsilon}
\def\x{\xi}
\def\z{\zeta}
\def\D{\Delta}
\def\F{\Phi}
\def\G{\Gamma}
\def\L{\Lambda}
\def\O{\Omega}
\def\P{\Pi}
\def\Q{\Theta}
\def\S{\Sigma}
\def\U{\Upsilon}
\def\X{\Xi}
\def\delb{\bar\del}

\title{Non-spherical collapse in AdS  and  \\ Early Thermalization in RHIC }

\author[a]{ Eunseok Oh}
\author[a]{and Sang-Jin Sin}
\affiliation[a]{Department of Physics, Hanyang Univ.  Seoul 133-791}
\emailAdd{lspk.lpg@gmail.com}
\emailAdd{sjsin@hanyang.ac.kr}

\abstract{In the flat space, non-spherical shells collapse to give globular cluster after many oscillations. We show that in anti de sitter space,   they    form  black holes in one dynamical time. 
We propose that  this is  the mechanism of early  thermalization in strong quark-gluon plasma in gravity dual. 
This is  traced back to the  
 a remarkable property of AdS : the period in radial motion is amplitude independent  in spite of the  NON-linearity of the equation of motion.
We investigate the interaction effect numerically and observe the same qualitative behavior for the   attractive forces. For repulsive interactions, particles halt at a 
small but finite radius for long time due to the specific structure of the 
bulk AdS propagator. It helps hair creation  in  the AdS black hole.     
 }


\maketitle

\section{Introduction}
\label{sec:Introduction}

One of the mysteries of RHIC experiment is the early thermalization\cite{Heinz:2001xi}. 
The fireball made from the collision seems to reach equilibrium in 1fm/c, which is   comparable time for gold ions  to pass each other. 
This is attributed to the strongly interacting nature of the system. 
 Since it is hard to understand this phenomena in perturbative field theory,   it is natural to ask if a dual formulation where  particles are weakly coupled  is helpful. According to the gauge/gravity duality  \cite{Maldacena:1997re,Gubser:1998bc,Witten:1998qj},   thermalization is   dual to the black hole formation in Anti de Sitter(AdS) space.  In a previous paper of one of the authors \cite{Shuryak:2005ia}, mapping 
the entire   process of RHIC experiment   to the dual gravity language was tried. Our question in this formulation  is 
"Why is the black hole formation 
time of a generic gravitational collapse in AdS space of the order
of the passing time of the two ions?" 
In other words, 
"Why does generic gravitational collapse produce black hole in one falling time in AdS space?"  

For the collapse of a spherical shell, it is less surprising that the final result is the black hole. However, for non-spherical shell, the result is far from a black hole in flat space, where a globular cluster rather than a black hole
is formed.  Recent  studies on scalar field collapse in AdS space 
\cite{Bhattacharyya:2009uu,Bizon:2011gg,Garfinkle:2011hm,Garfinkle:2011tc,Buchel:2012uh} shows that even for the spherical symmetric collapse, black hole is made only after repeated  reflections from the boundary.  Therefore such results  have some distance from  `thermalization in one passing time'.  The natural explanation for the experiment should be such that the black hole is formed 
in one falling time for any generic initial configuration.

Also, showing the black hole formation by numerical analysis does not give  an `understanding' why it is formed so easily in AdS  but not in flat space.
While some analytic discussion is provided in \cite{Bizon:2011gg}, 
it is not  very clear  to us what the physical  mechanism of inevitable  black hole formation is. Confining or Box nature of AdS space does NOT explain the easy formation of black hole. 

 In this paper, we will study the collapse of non-spherical shell
 made of dusts,  
 and show that it forms a black hole in one finite falling time 
due to a specific property of  the AdS.  
 We treat the shell as a collection of dust  particles, which is   
   very different from the scalar field collapse where it is assumed that matter is in a state of coherent condensation. Only when particle's wave functions are overlapping, one can justify  treating many particles  in terms of condensation wave function or the scalar field configuration.  
 
 In the gauge/gravity duality one considers the limit where string is much smaller than the AdS radius. In this limit the wave nature as well as the stringy nature  is suppressed. For the many particle system in RHIC,   a scalar field configuration  is not the proper dual configuration to the fireball. 
Therefore we consider the shell as a collection of the   particles 
in the global AdS. We will show that particles arrive at the center simultaneously regardless of  their masses and initial positions  as far as 
they start from  static configurations. 
 This means that any non-spherical shell in AdS 
 space becomes more spherical shell as it falls, and all parts of the shell reach the center simultaneously. We call this property as synchronization effect of AdS. 
 After such shell pass the 
`would be horizon', black hole forms and particle's motion can not and need not 
be traced if we take into account the back reaction of the gravity.  
It proves that any cloud falls to become a  black hole in AdS space.  

After this we consider the phenomena in the Poincare patch where 
 AdS/CFT is manifest. We will see that there will be a finite time which is less than 
 one falling time at which all the particle are inside the radius of apparent horizon as far as there is no IR cut off  in the geometry.  
We will first show this for   particles without inter-particle interaction; then we will show that 
the same thing is true even in the presence of the interaction. 
In gauge/gravity duality, the 
dynamics of gluon exchange is treated  by gravity background in leading 
order of 1/N expansion. So here neglecting inter-particle interaction means treating particle interaction only  in leading  1/N expansion of gauge theory. 
 
 Even in the simplest case of free fall along the radial direction, 
  the equation of the  motion (EOM) is that of a non-linear oscillator.  
Nevertheless its period is independent of the amplitude as if it is a simple harmonic oscillator(SHO). This is a remarkable property of  AdS spacetime. We will de-mystify this phenomena by finding  a non-linear mapping that transforms the EOM of the falling into that of SHO. 
In fact, the idea of Synchronization was first formulated in ref. \cite{Shuryak:2005ia} where it was observed that 
the radial motion is that of SHO in the proper time and in Poincare coordinate.  However each particle has its own proper time and therefore  it could not  be  claimed definitely that 
particles arrive at the center  at the same time based on the "same falling proper time".  Also in that paper a geometry with 
IR brane describing the confining phase was used, 
which can raise the question: how a system in confining phase can   
go to a deconfining phase by cooling? 
 As it was argued in \cite{Shuryak:2005ia}, the falling is dual to the expansion which cause the system to cool down.  
One of the purpose of this paper is to improve these difficulties.

  %

\section{Collapse in global AdS$_5$ without inter-particle interaction}
\label{sec:Proca}
 
To consider the collapse of  non-spherical shell which consists of 
non-interacting particles, we need to look at the motion of individual particles. 
For  $AdS_5$ with spherical boundary the metric is given by 
\be
ds^2=-(1+r^2/R^2)c^2dt^2+r^2d\Omega^2 + \frac{dr^2}{1+r^2/R^2}
\ee
It is well known that the falling time from the boundary to the center 
following the null geodesic is $\pi R/2c$. However we will see 
that even the massive particle starting from arbitrary position 
with zero radial velocity will arrive at the center after the same time. 
Notice that the time for the light starting from the arbitrary position 
to arrive at the center will not be the same.  

The equation of motion is given by the action 
\be
S=- m\int \sqrt{- g_{\mu\nu} \dot{x^\mu} \dot{x^\nu}} dt, \;{\rm with} \;\; {\dot x}=\frac{dx}{dt} . 
\ee
For simplicity,  we first consider the radial motion and set $R=1, c=1$.
The energy conservation can be written as 
\be
 \frac{m(1+r^2)}{\sqrt{1+r^2 -{\dot r}^2/(1+r^2)}}=E.
 \label{1st-int}
 \ee
 It is easy to see that the system describes a non-linear oscillator. 
 Notice that the dot in the above equation is derivative in $t$ not in the proper time unlike ref.  \cite{Shuryak:2005ia}.  
 If we assume that the particle starts with zero radial velocity from the 
 initial radial position $r_0$, then 
 \be E=m \sqrt{1+r_0^2},
 \ee 
 which establishes a dictionary between the total energy and the initial radial coordinate. 
Introducing   $v_c$   by $v_c=r_0/\sqrt{1+r_0^2}$, we have 
 \be 
 E = \frac{m}{\sqrt{1-v_c^2}}.\ee
 Its velocity in the radial direction starts with 0 and   become  $v_c$ 
 when it arrive at the center. 
Interestingly, we can  find the exact solution of the equation of motion: 
 \be
 r=\frac{v_c\cos t}{\sqrt{1-v_c^2\cos^2  t}}.
 \ee
The remarkable property of this solution is that the period of the motion is 
$2\pi$, independent of the original position $r_0$, as if it is a simple harmonic oscillator.
Restoring  the scale parameters $R, c$ by 
   $r,t,v \to r/R, t/R,v/c$   the falling time is 
\be
T_{fall}= \frac{\pi  }{2} \frac{R }{c}.
\ee

This means that arbitrary set of particles, falling in AdS will form a black hole regardless of the initial position, provided,  
(i) all the particles start with zero initial  velocities, and 
(ii) Interaction  between the particles are negligible.  
See figure 1. 
\begin{figure}[t]
\centering
\includegraphics[width=6cm]{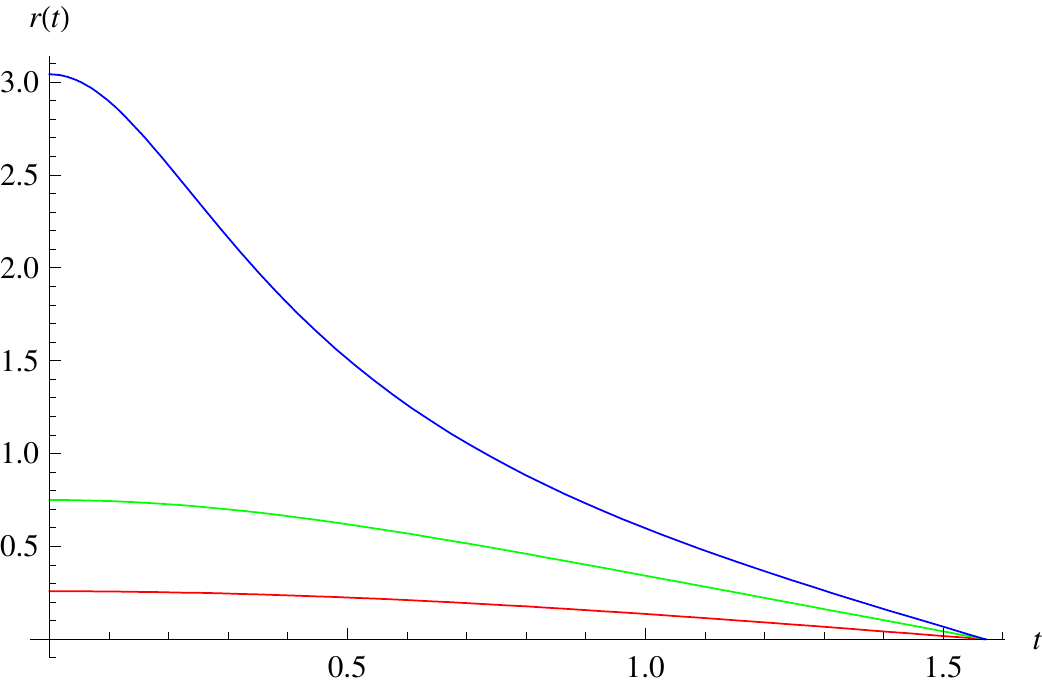}
\includegraphics[width=8cm]{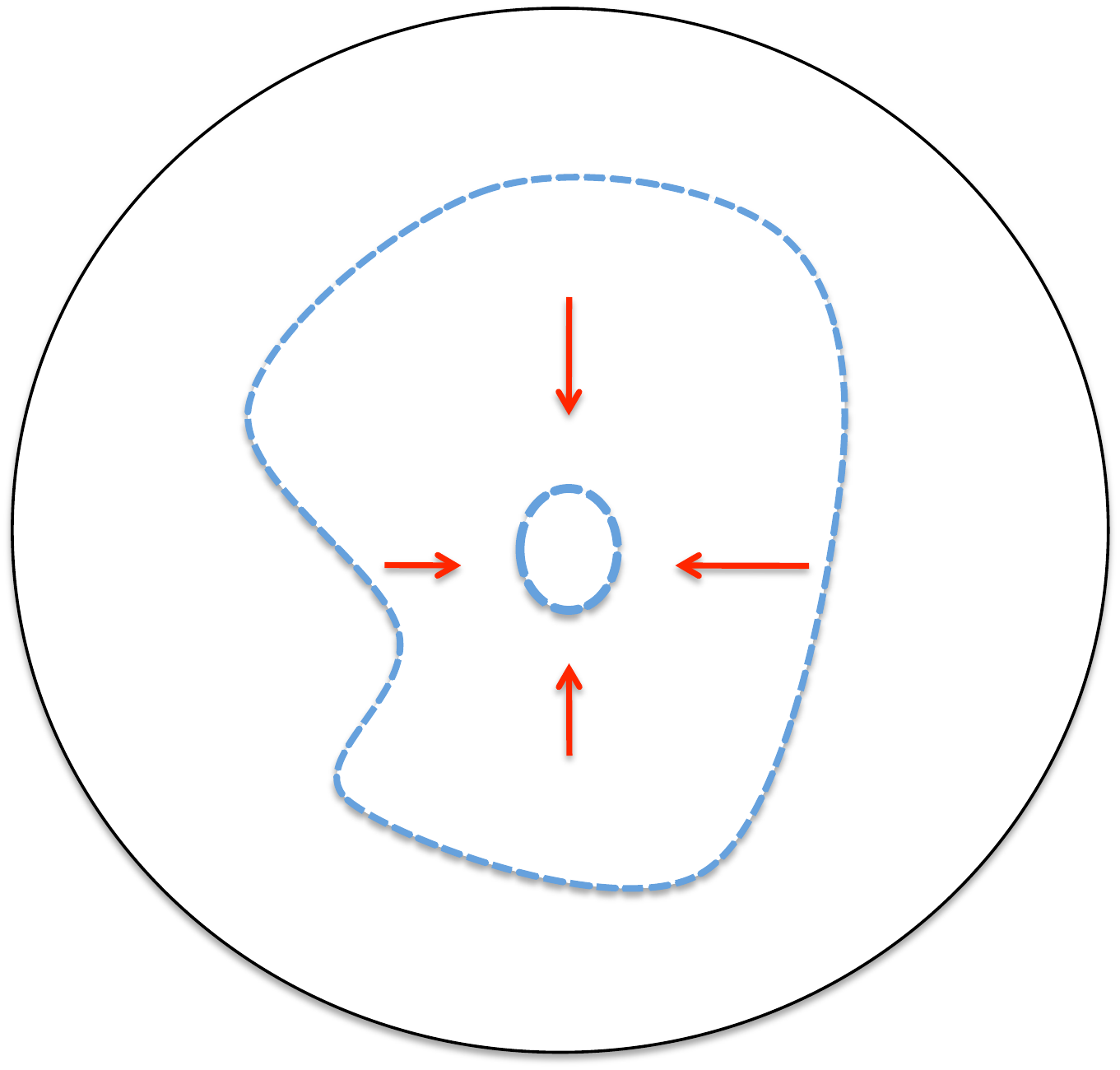}
\caption{Left: Free falling of a particle in AdS for various initial positions. Falling time is independent of the initial position;    Right: Collapse  of a shell with arbitrary shape in AdS space. 
The whole shell will reach at the center simultaneously.  
$T_{fall}= \frac{\pi}{2} \frac{R}{c}$. } 
\label{fig:fall}
\end{figure}

How do we justify or relax  the conditions listed above? 

\begin{enumerate}
\item
In AdS/CFT correspondence,  
the radial direction is the dual of the energy scale.
If two gold ions collided in RHIC and created particles of energy $E_i$ and  mass $m_i$,     
$i=1,\cdots, N$ inside the fireball, then the holographic image of such 
fireball is the 
particles in AdS space at the position 
\be
r_{0i}=\sqrt{(E_i/m_i)^2-1}.
\ee
Particles created in the fireball have velocities in $x^i$ directions but not in radial direction. Therefore their dual image particles do not either. 
This justifies  the first issue listed above. 
\item
In gauge/gravity dual, the gluon dynamics is replaced by the gravitational 
background and the leading order 
inter-particle gluonic interaction in the many body quark-gluon plasma 
is approximated as the particle motions in a fixed gravitational background. 
  Interparticle interaction in AdS is necessary only
to take care of the non-gluon interaction or non-leading order interaction in ${\cal O}(1/N_c)$; therefore, such inter particle interaction in AdS bulk should be absent or very weak in the large $N_c$ theory. 
\end{enumerate} 
 
To de-mystify the amplitude independence of the 
period of the above  nonlinear oscillators, we can actually map 
the above non-linear oscillator equation to that of the harmonic oscillator. 
If we define $\eta$ by 
\be
r=\frac{\eta}{\sqrt{1-\eta^2}},
\ee
the eq.(\ref{1st-int}) can be mapped to the  harmonic oscillator;
\be
{\dot \eta}^2 + \eta^2=\eta_0^2:=1-\left(\frac{m}{E}\right)^2,
\ee
whose solution is $\eta=\eta_0 \cos t$. 
In this coordinate, the metric becomes 
\be
ds^2=  \frac1{1-\eta^2}\left( -  dt^2+\eta^2 d\Omega^2\right) + 
\left(   \frac{ d\eta}{1-\eta^2}  \right)^2.
\ee
We can also define $\eta=\tanh \rho$ to get $r=\sinh\rho $ 
to get the metric in the global coordinate:
\be
ds^2=-\cosh^2\rho dt^2+d\rho^2 +\sinh^2\rho d\Omega^2 .
\ee

The point is that once the particles pass the would-be horizon of the system, there is no turning back. 
Apparent horizon formation is inevitable consequence of the synchronized falling. 

\section{Flat boundary case}
Now we work in Poincare patch where    gauge/duality is well established. The metric is given by 
\be
ds^2= \frac{r^2}{R^2}(-c^2dt^2+ d{\bf x}^2) + R^2 \frac{dr^2}{ r^2}
\ee
 and it has flat boundary. 
 The  integrated  equation of motion is 
 \be
 \frac{mr^2}{\sqrt{r^2 -{\dot r}^2/r^2}}=E,
 \ee
and its solution is \cite{Shuryak:2005ia}  
\be
r=  \frac \epsilon{ \sqrt{1+ (\epsilon t)^2}} .
\ee
This is not a periodic solution. It show that starting from $r_0=E/m=\epsilon$, it takes infinite time to reach the center. 
However, the large time behavior, $r\sim 1/t$,  is  independent of the initial   height $r_0=\epsilon$. 
This is a manifestation of the synchronization effect in this metric: two particles with different initial height fall and get closer, which is enough to argue the canonical formation of the black hole in this metric.   
This effect was observed in \cite{Shuryak:2005ia} 
as a `tendency' of time focusing. 
However,  due to the 
infinite falling time in Poincare coordinate time, 
it was not recognized there that synchronization is the 
exact property of the AdS space, partly because IR cut off  was introduced
in the work.   
\footnote{The  concept  of synchronization was initiated   in \cite{Shuryak:2005ia} 
   due to the fact that the radial  equation of motion in the Poincare patch was 
found to be simple harmonic oscillator in the proper time. However, 
each particle has its own proper time and  it  could  not be concluded that exact synchrozation is the property of AdS from this either.
Notice that the SHO discovered in this paper is in real time not proper time. 
}
Later, we will show that  introducing the interaction does not change the synchronization property of AdS. 

\subsection{Effect of the initial velocity}
Here we study the effect of the initial velocity along the space time direction.  Suppose the particle is in motion along $x$ direction. Then two first integrals are 
\be
 \frac{mr^2}{\sqrt{r^2 (1-{\dot x}^2)-{\dot r}^2/r^2}}=E, \\
  \frac{mr^2 {\dot x}}{\sqrt{r^2 (1-{\dot x}^2)-{\dot r}^2/r^2}}=p, 
 \label{1st-int4}
 \ee
 which can be called as the energy and momentum respectively. 
If we set $V=p/E$,  
\be
r=  \frac {\epsilon (1-V^2) }{ \sqrt{1+ (\epsilon(1-V^2) t)^2}} .
\ee
we get ${\dot x}=V$.
Remarkably the large time behavior of the radial position is 
independent of all of the initial conditions $m, E, p$. 
Therefore we can say that the time focusing effect is perfect even in the presence of the motion along the collisional direction.

\section{The effect of the  Interaction}
 So far, we discussed the particles falling  without interaction. 
 We now discuss the effect of it  in Poincare coordinate where the metric is  
\be
ds^2=\frac1{x_0^2}(dx_0^2+d{x^\mu} d{x_\mu})
, {\hbox{ with }} x_0=1/r\ee
The scalar   propagator for particle with mass $m^2=\Delta(\Delta-d)$  in the AdS$_{d+1}$ is given in \cite{D'Hoker:1998mz} 
and it is given by
\be
G\sim \left( \frac{1}{u(2+u)} \right)^\Delta 
\hbox{ with } u=\frac{{(x-y)^M (x-y)_M}}{2x_0y_0}
\ee

Since $u$ is nonnegative and we are interested in the  most singular contribution, we will neglect the factor $u+2$. 
The  Newtonian potential in AdS can be derived from this to give 
\bea
V(\{x_i,y_i\})&=&\int \int d^5x d^5y J(x)G(x,y)J(y),  \\
&=& G_N \sum_{i<j} \int dt 
\frac{(x_{i0}x_{j0})^{2\Delta}}{ \big( |x_i(t)-x_j(t)|^2+|x_{i0}(t)-x_{j0}(t)|^2 \big)^{\Delta -1/2} }
\eea
where $J(x)=\sum_i \delta^4(x^A-x^A_i(t)), A=0,1,2,3 .$
Since the initial velocities are only along the $x^\mu$ direction, 
we expect that the attractive interaction will only enhance the focusing effect in radial motion. Indeed we can verify this numerically, 
using the equation of motion 
derived from the   Lagrangian 
\be
S=- m\int \sqrt{- g_{\mu\nu} \dot{x^\mu} \dot{x^\nu}} dt, \;
+ V(\{x_i,y_i\})
\ee
 In the numerical calculation we used $G_N=1, \Delta=3/2$ for simplicity. 
 
In figure 2, falling of a few particles with  some horizontal  initial velocities starting from   different heights are drawn.
The calculation is done by the mathematica.  The vertical lines are the falling trajectories and 
the horizontal dashed lines indicate the equal time slices.  
As time goes on,  it is manifest that the radial positions converge. 
\begin{figure}[t]
\centering
\includegraphics[width=2.5cm]{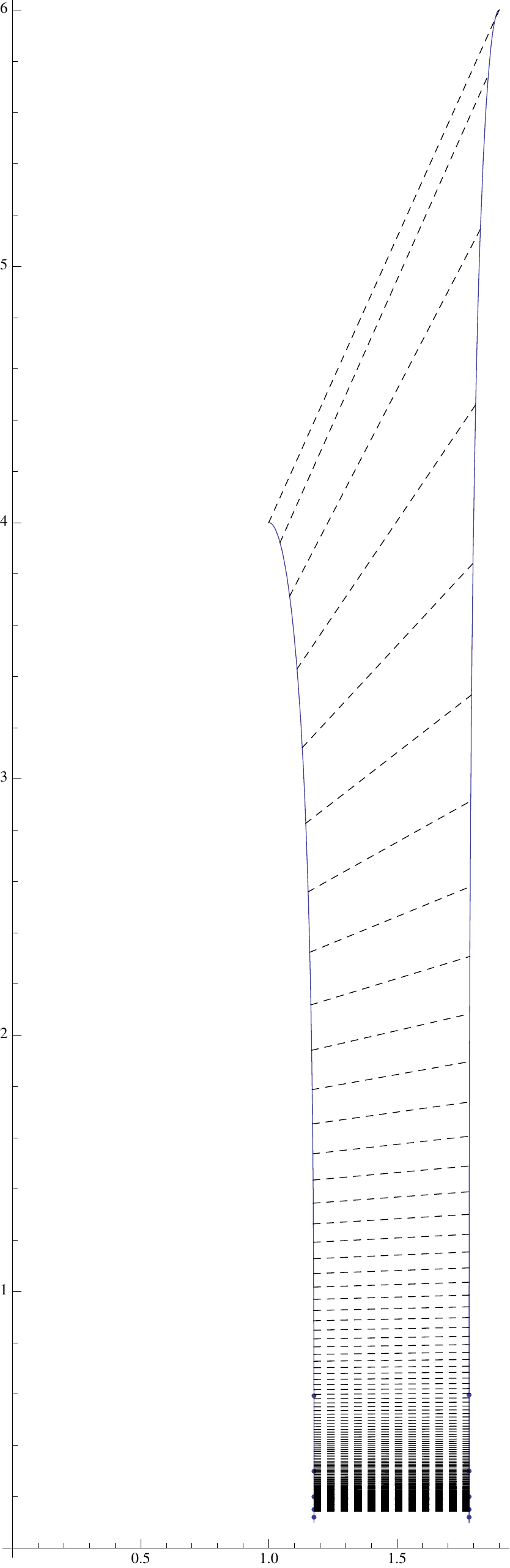}
\includegraphics[width=2.5cm]{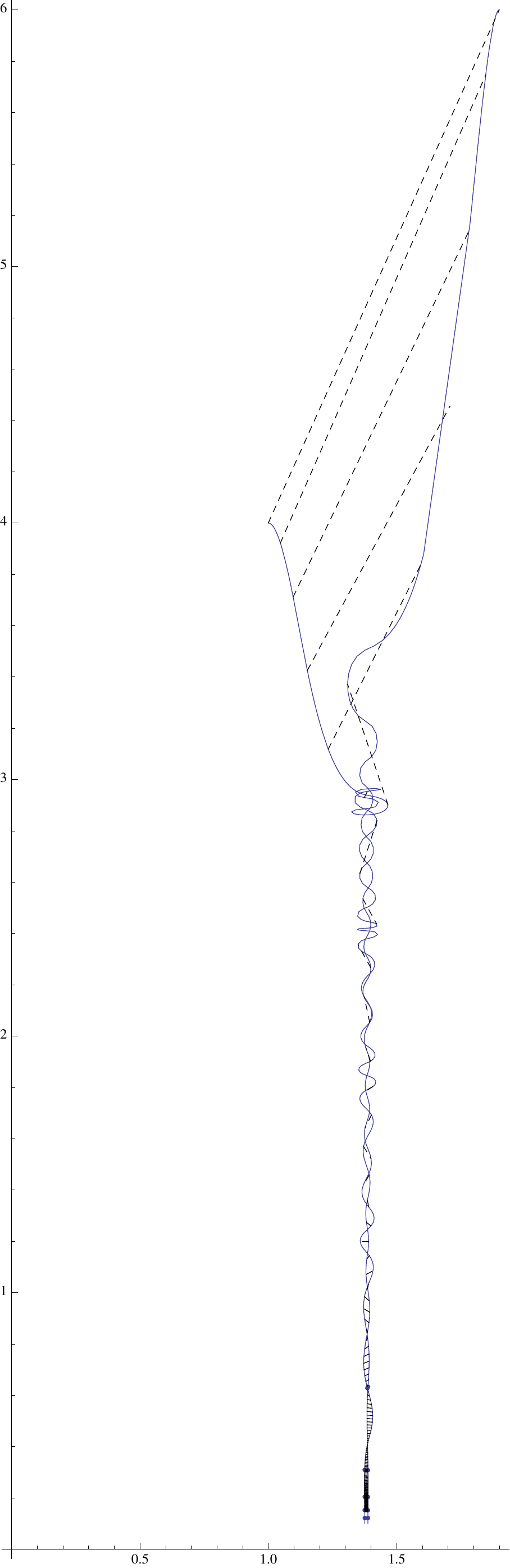}
\includegraphics[width=8cm]{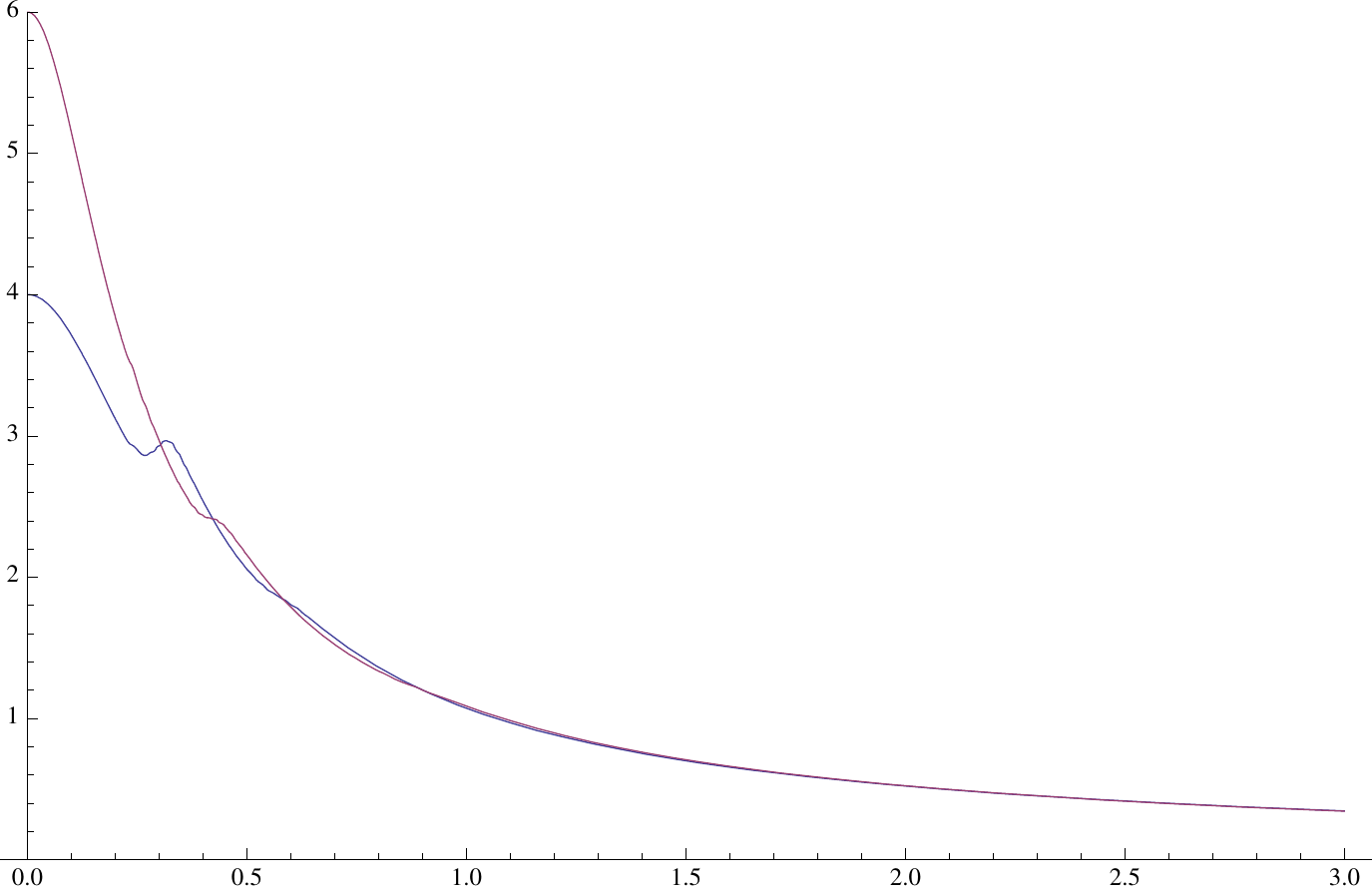}
\caption{Falling in AdS with and initial velocity. : (Left) without inter-particle interaction. This is the flat-boundary analogue of Figure 1.
(Middle) with interaction.   $r$ v.s x, x-axis is one of the space direction. (Right) radius as a funcntion of time. 
The synchronization  effect (equalizing the radial position) is manifest here. 
} 
\label{fig:fig2}
\end{figure}
In figure 3,  we consider  what happens if inter-particle interactions are repulsive.
This is possible if particles carry extra charges. 
   Interestingly, we see that falling is halted for some moments at the certain radial region 
and then proceeds to the  black hole formation.
We can understand such behavior  from the structure of the propagator (4.2) : In the deep IR region, $u\to 0$  for any finitely distant two points, therefore strong repulsion is effective there although particles are separated enough. 
We speculate that this can be the mechanism of the formation of the gravitational hair which is observed in the theory of holographic superconductor.

In figure 4, we show what happens if three particles  collide with inter-particle attraction and repulsion. 
As we can see, whatever is the situation, radial positions of particles converges. Therefore we conclude that such synchronization effect is not destroyed by the interaction effect, especially if the interaction is attractive.   
For the repulsive case, particles halts at small but finite radius for long time due to the specific structure of the 
propagator: namely, $x_0y_0$ factor in $1/u$. See (4.2) and (4.4).  We believe that this is responsible for the 
presence of the hair in the AdS space. 

\begin{figure}[t]
\centering
\includegraphics[width=6cm]{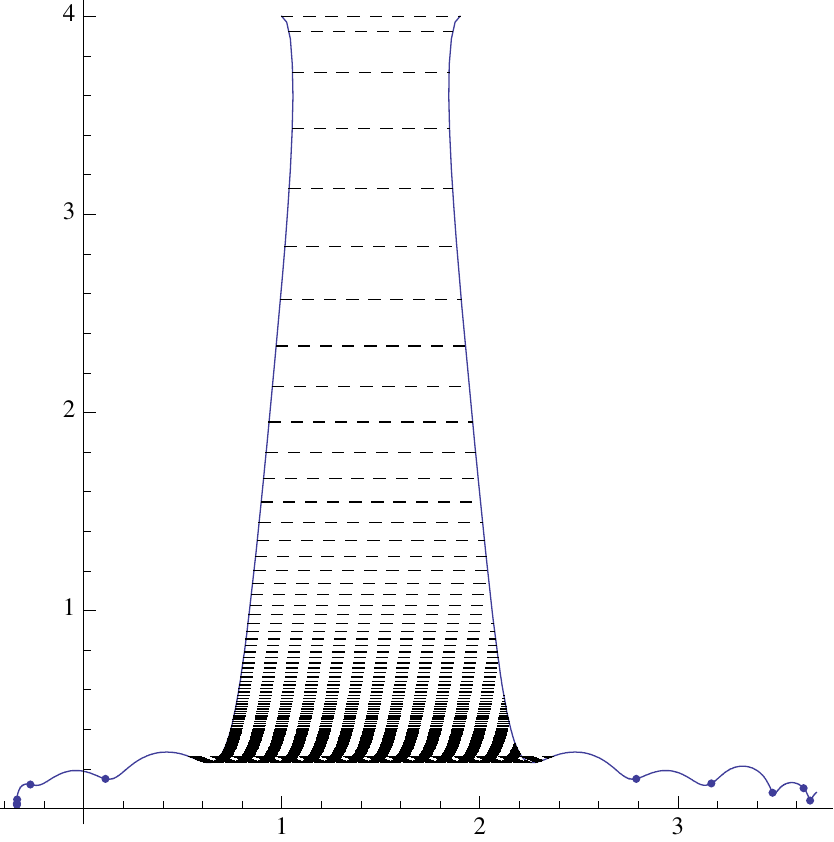}
\includegraphics[width=4cm]{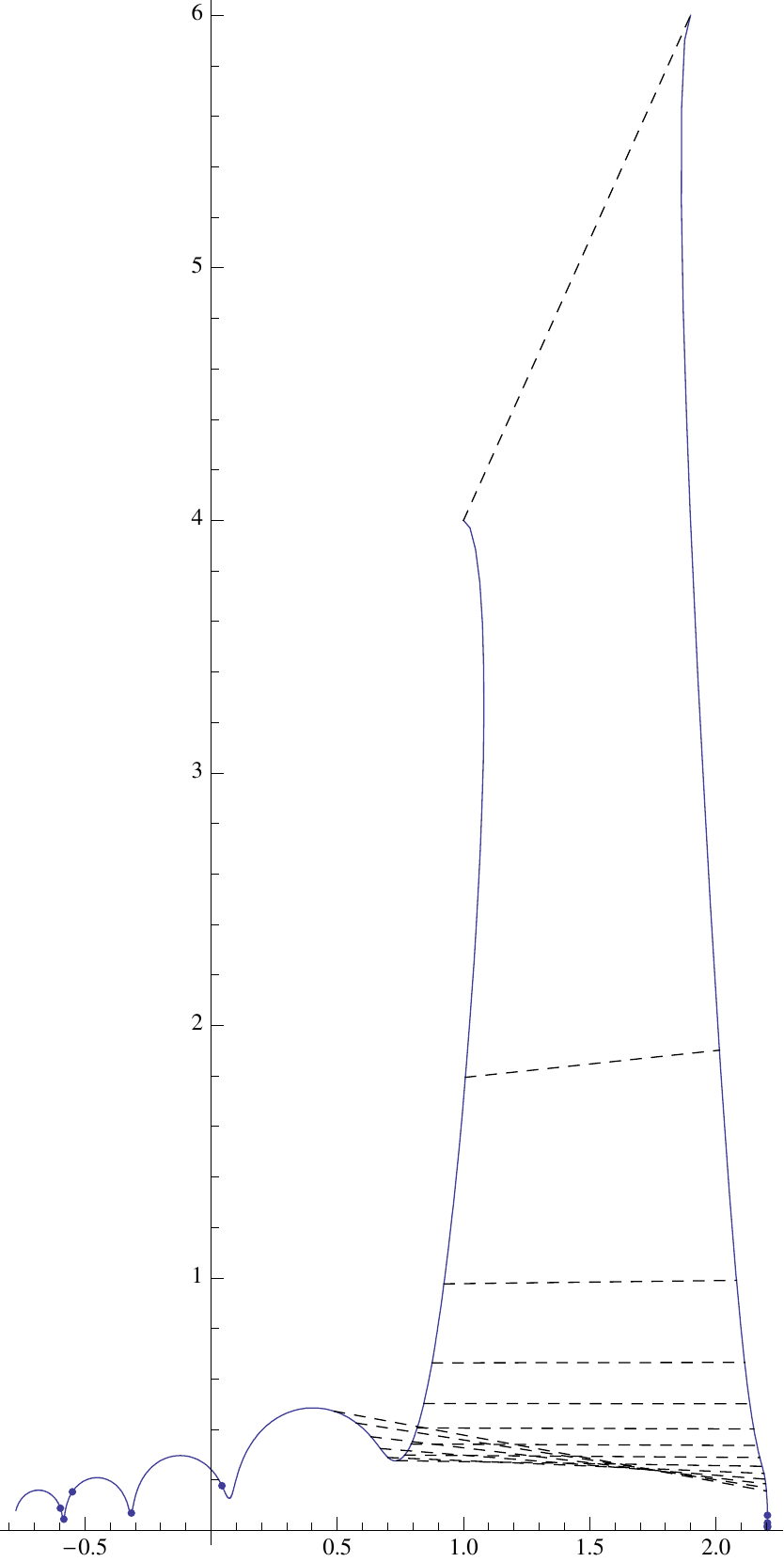}
\caption{$r$ v.s $x$. 
 Falling in AdS with repulsive interaction. L) start from the same height.  R) start from the different height. 
 } 
\label{fig:fig3}
\end{figure}

\begin{figure}[t]
\centering
\includegraphics[width=6cm]{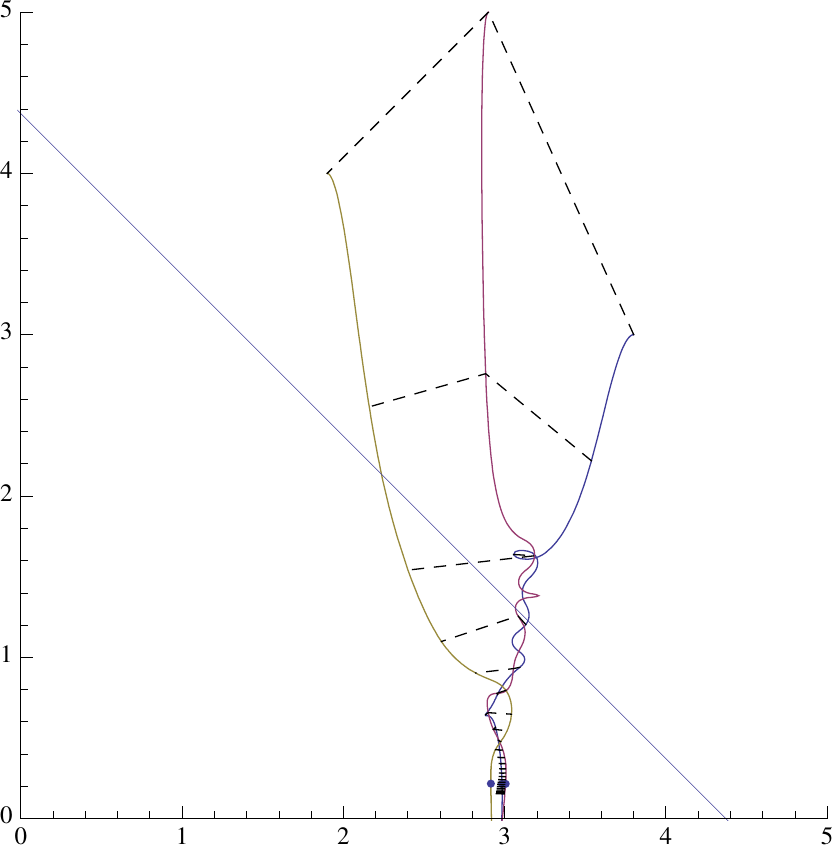}
\includegraphics[width=6cm]{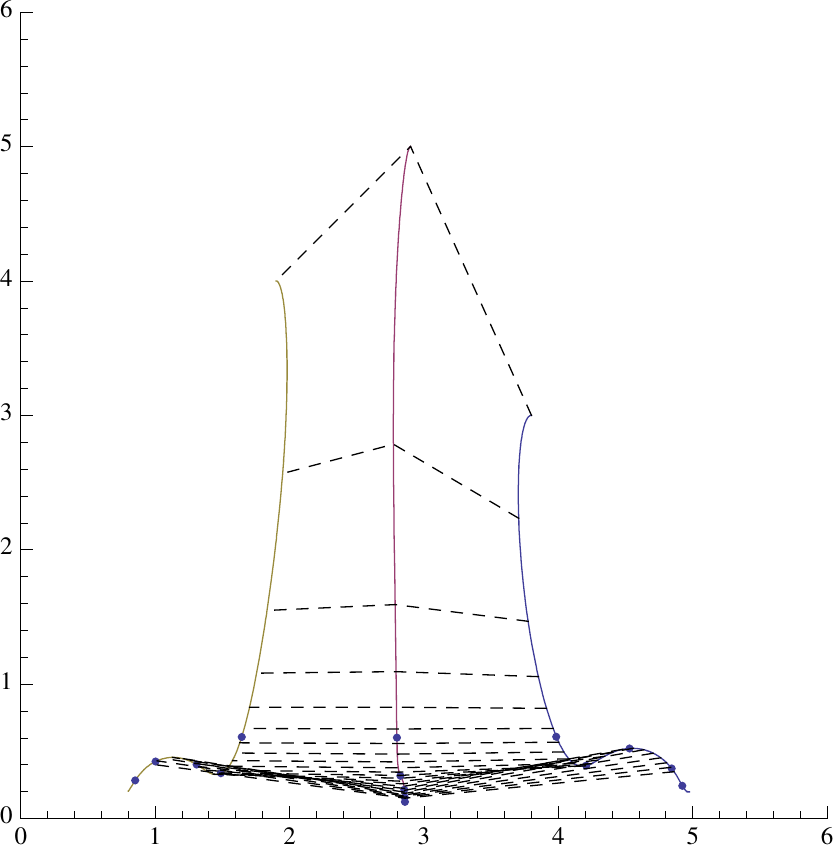}
\caption{Three particles colliding in 5 dimension. L) with attractive interaction. R) with repulsion.  All the particle arrive at the center simultaneously even in the presence of the interaction. } 
\label{fig:fig4}
\end{figure}

\section{Conclusion}

In this paper we demonstrated that 
arbitrary shape of shell in AdS falls and form a black hole. 
The physical mechanism is  the synchronized 
falling  which is the characteristic property of AdS. 
While individual particle's motion is oscillation, 
many particle's motion should be terminated by the black hole
formation. Once the black hole is formed, there will be no more oscillation. 
The details of the stabilization of the system through  losing the potential energy should be  worked out  by considering the back reaction of the metric, which is not the scope of this work. 



 
 The reason for several oscillation before black hole formation in the study of the spherical scalar field collapse 
 is due to the wave nature of the  scalar field:  as the field configuration collapses, uncertainty principle  activates the kinetic term \cite{Sin:1992bg}, which generates pressure and causes the bounce. 
 It can be also attributed to the effective repulsion between the 
 scalar particles as we have seen in the section 4. 

The future work is to find the effect of the extra interactions and to find the time dependent solution of Einstein equation that describes the formation of the black hole. Large time approximate solution was proposed in \cite{Janik:2005zt} and 
further developed in 
  \cite{Nakamura:2006ih,Sin:2006pv}.  
However, finding exact Einstein gravity solution  is non-trivial even for the   spherically symmetric shell and we will postpone it to future work.

  \acknowledgments
 This work was supported by Mid-career Researcher Program through NRF grant No. 201300000001340. 
 It is also partly supported by the  NRF  grant  through the SRC program  CQUeST with grant number 2005-0049409. 


%

\begin{thebibliography}{10}

\bibitem{Heinz:2001xi} 
  U.~W.~Heinz and P.~F.~Kolb,
  Nucl.\ Phys.\ A {\bf 702}, 269 (2002)
  [hep-ph/0111075].
  

\bibitem{Maldacena:1997re}
J.~M. Maldacena, {\it {The Large N limit of superconformal field theories and
  supergravity}},  {\em Adv.Theor.Math.Phys.} {\bf 2} (1998) 231--252,
  [\href{http://xxx.lanl.gov/abs/hep-th/9711200}{{\tt hep-th/9711200}}].

\bibitem{Gubser:1998bc}
S.~Gubser, I.~R. Klebanov, and A.~M. Polyakov, {\it {Gauge theory correlators
  from noncritical string theory}},  {\em Phys.Lett.} {\bf B428} (1998)
  105--114, [\href{http://xxx.lanl.gov/abs/hep-th/9802109}{{\tt
  hep-th/9802109}}].

\bibitem{Witten:1998qj}
E.~Witten, {\it {Anti-de Sitter space and holography}},  {\em
  Adv.Theor.Math.Phys.} {\bf 2} (1998) 253--291,
  [\href{http://xxx.lanl.gov/abs/hep-th/9802150}{{\tt hep-th/9802150}}].
  
  
 \bibitem{Shuryak:2005ia} 
  E.~Shuryak, S.~-J.~Sin and I.~Zahed,
  J.\ Korean Phys.\ Soc.\  {\bf 50}, 384 (2007)
  [hep-th/0511199].
 
  
\bibitem{Bhattacharyya:2009uu} 
  S.~Bhattacharyya and S.~Minwalla,
  JHEP {\bf 0909}, 034 (2009)
  [arXiv:0904.0464 [hep-th]].


\bibitem{Bizon:2011gg} 
  P.~Bizon and A.~Rostworowski,
  Phys.\ Rev.\ Lett.\  {\bf 107}, 031102 (2011)
  [arXiv:1104.3702 [gr-qc]].
  
  \bibitem{Garfinkle:2011hm} 
  D.~Garfinkle and L.~A.~Pando Zayas,
  Phys.\ Rev.\ D {\bf 84}, 066006 (2011)
  [arXiv:1106.2339 [hep-th]].
  
  \bibitem{Garfinkle:2011tc} 
  D.~Garfinkle, L.~A.~Pando Zayas and D.~Reichmann,
  JHEP {\bf 1202}, 119 (2012)
  [arXiv:1110.5823 [hep-th]].
  
\bibitem{Buchel:2012uh} 
  A.~Buchel, L.~Lehner and S.~L.~Liebling,
  Phys.\ Rev.\ D {\bf 86}, 123011 (2012)
  [arXiv:1210.0890 [gr-qc]].
  
  \bibitem{Sin:1992bg} 
  S.~-J.~Sin,
  Phys.\ Rev.\ D {\bf 50}, 3650 (1994)
  [hep-ph/9205208].
   
  \bibitem{D'Hoker:1998mz} 
  E.~D'Hoker and D.~Z.~Freedman,
  Nucl.\ Phys.\ B {\bf 550}, 261 (1999)
  [hep-th/9811257].
  \bibitem{D'Hoker:1999jc} 
  E.~D'Hoker, D.~Z.~Freedman, S.~D.~Mathur, A.~Matusis and L.~Rastelli,
  Nucl.\ Phys.\ B {\bf 562}, 330 (1999)
  [hep-th/9902042].
 

 \bibitem{Janik:2005zt} 
  R.~A.~Janik and R.~B.~Peschanski,
  Phys.\ Rev.\ D {\bf 73}, 045013 (2006)
  [hep-th/0512162].
  
  \bibitem{Nakamura:2006ih} 
  S.~Nakamura and S.~-J.~Sin,
  JHEP {\bf 0609}, 020 (2006)
  [hep-th/0607123].
  
\bibitem{Sin:2006pv} 
  S.~-J.~Sin, S.~Nakamura and S.~P.~Kim,
  JHEP {\bf 0612}, 075 (2006)
  [hep-th/0610113].
  
\end{thebibliography}
\end{document}